\documentclass[preprint,12pt]{elsarticle}
\usepackage{amssymb}
\usepackage{amsmath}

\begin{document}
\begin{frontmatter}

\title{Influence of growth parameters on the superconducting transition temperature in granular aluminum films}

\author{Aniruddha Deshpande}
\author{Jan Pusskeiler}
\author{Christian Prange}
\author{Uwe Rogge}
\author{Martin Dressel}
\author{Marc Scheffler}
\ead{marc.scheffler@pi1.uni-stuttgart.de}
\affiliation{organization={1.\ Physikalisches Institut, Universit{\"a}t Stuttgart},
            addressline={70569 Stuttgart},
             country={Germany}}

\begin{abstract}
This study investigates the influence of various growth parameters on normal-state resistivity and superconducting transition temperature $T_{c}$ of granular aluminum films. Specifically, we focus on the effects of oxygen flow and aluminum evaporation rate during the growth process conducted at different substrate temperatures, from 300~K down to 25~K. We report systematic correlations between the growth conditions, the normal-state resistivity, and $T_{c}$. Furthermore our findings provide insights into optimizing the superconducting characteristics of granular aluminum.
\end{abstract}


\end{frontmatter}

\section{Introduction}\label{sec1}

Granular superconductors, composed of small metallic grains dispersed in an insulating matrix, present a fascinating intersection of superconductivity and disorder which can also be imagined as grains of homogeneous superconductor separated by insulating barriers or Josephson junctions \cite{Abeles1966, Abeles1967, Parmenter1967, Cohen1968, Deutscher2021}. These materials are of significant interest in condensed matter physics due to their unique ability to exhibit both superconducting and insulating behaviors, depending on their microscopic structure and specific environmental conditions and parameters at play \cite{Kunchur1987, Bachar2023}. The physical properties of granular superconductors are highly sensitive to the microstructure, particularly the size, distribution, and coupling between the grains \cite{Deutscher1973_JLTP}. This sensitivity allows researchers to explore a wide range of quantum phenomena, including the superconductor-insulator transition (SIT), where the material can switch from a superconducting state to an insulating state as a function of resistivity or external parameters such as magnetic field or temperature \cite{Jaeger1989, Frydman2002, Belobodorov2007, Grenet2007, Bachar2015_PRB, LevyBertrand2019}. Granular superconductors serve as model systems for studying the effects of disorder in superconductivity, providing insights that are relevant for understanding more complex systems \cite{Mayoh2014}. The interplay between the superconducting phases of the grains, their size, and the coupling between them, along with the insulating matrix, introduces a rich variety of behaviors that make granular superconductors not only a subject of fundamental scientific curiosity but also a potential platform for novel technological applications, particularly in cryogenic sensors and quantum computing \cite{Valenti2019, LevyBertrand2021, Winkel2020_PRA, Winkel2020_PRX, Gruenhaupt2018}.

Granular aluminum consists of pure aluminum grains with a diameter of about 2-3~nm embedded in a matrix of amorphous aluminum oxide (Al$_{2}$O$_{x}$), which self-assemble during the deposition of aluminum in an oxygen-rich atmosphere \cite{Deutscher1973_JLTP}. During the growth process, oxygen in the chamber controls the thickness of the oxide barrier between the grains. 
This thickness can range from a thin monolayer to a few nanometers, depending on the oxygen concentration, which in turn, influences the normal-state resistivity \cite{Ziemann1978}. It is one of the most extensively studied disordered superconductors, especially due to its high degree of low-temperature kinetic inductance up to several nH per square \cite{Gruenhaupt2018, Kamenov2020} and robustness against external magnetic fields \cite{Abeles1967, Cohen1968, Chui1981, Borisov2020}. 
These properties make granular aluminum a versatile material for superconducting quantum circuits \cite{Winkel2020_PRX, Gruenhaupt2018, Kamenov2020, Maleeva2018, Schoen2020, Rieger2023}, parametric amplifiers \cite{Eom2012,Zapata2024}, kinetic inductance detectors (KIDs) \cite{Valenti2019, LevyBertrand2021, Day2003}, nanoSQUIDs \cite{Avraham2023}, and other quantum devices \cite{EsmaeilZadeh2021, Rotzinger2017, Irwin2010, Kurter2010}. 
Furthermore, the frequency-dependent properties of granular aluminum exhibit characteristic GHz electrodynamics due to high kinetic inductance and signatures of collective modes at higher frequencies \cite{LevyBertrand2019, Maleeva2018, Steinberg2008, Bachar2015_JLTP, Pracht2016, Pracht2017, Moshe2019}.
Granular aluminum is known for its conceptually relatively easy fabrication and the ability to tune the superconducting properties of the film by manipulating its microstructure during its growth process \cite{Deutscher2021, Bachar2023, Deshpande2025}. 
This tunability is achieved through precise control of growth parameters such as oxygen flow rate, aluminum evaporation rate, and substrate temperature, which collectively influence the formation of the granular structure and the degree of disorder within the film. 
The material exhibits a dome-like phase diagram in the superconducting regime for the superconducting transition temperature $T_{c}$ as a function of normal-state resistivity \cite{Bachar2023, Deutscher1973_JLTP, LevyBertrand2019, Valenti2019, Pracht2016, Moshe2020,  Deutscher1973_JVST}, similar to the doping-dependent phase diagram seen in oxide superconductors \cite{Caviglia2008, Varma2010, Shibauchi2014, Keimer2015, Thiemann2018, Gastiasoro2020}. 
In this superconducting regime, $T_{c}$ initially increases with resistivity due to increasing grain decoupling and thus stronger impact of enhanced superconductivity in individual grains \cite{Pracht2016, GarciaGarcia2007, Bose2014}, reaching a broad peak, before decreasing as the grains become even more decoupled and the decreasing phase coherence weakens macroscopic superconductivity at higher resistivities \cite{Pracht2016, BacharDissertation, Emery1995}. 

This study investigates the influence of various growth parameters on the normal-state resistivity and the superconducting transition temperature of granular aluminum films, focusing on the effects of oxygen flow, aluminum evaporation rate, and substrate temperature. Our report offers an overview of how growth conditions influence superconductivity, providing a starting point for researchers looking to optimize granular aluminum films for specific applications like quantum sensing,  quantum computing, and other low-temperature investigations.

\section{Experimental Methods}\label{sec2}

\begin{figure}[tb]
\centering
\includegraphics[width=12.5cm]{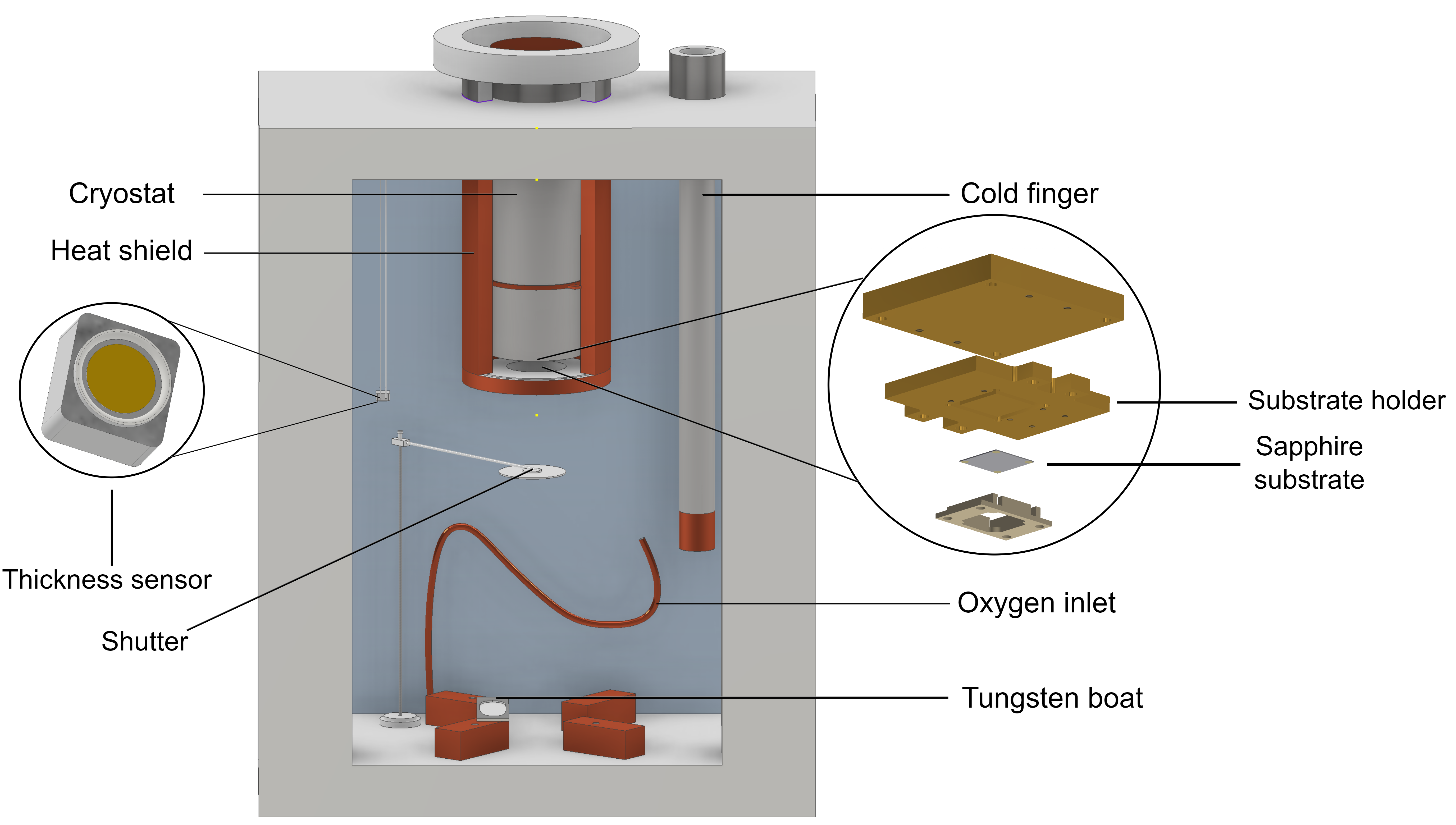}
\caption{A schematic drawing of the thermal evaporation chamber for growing superconducting granular aluminum thin films. Key components include the cryostat, heat shield, shutter, and oxygen inlet. Insets show detailed views of the quartz crystal thickness sensor (left) and the substrate holder assembly (right).}
\label{Fig:ChamberSchematic}
\end{figure}

The deposition system used in this study is shown schematically in Fig~\ref{Fig:ChamberSchematic}. The granular aluminum films were prepared by thermal evaporation of high-purity aluminum (99.98\%) from a tungsten boat onto R-plane sapphire substrates in the presence of oxygen. These substrates were mounted onto a brass substrate holder which was held at a distance of 28~cm from the tungsten boats. To enable the growth of thin films at lower temperatures, the substrate is coupled to the base of a cryogenic tank via an additional brass plate containing a temperature sensor and a heater for temperature stabilization. All the samples in the paper were grown on $10 \times 10$~mm$^2$ substrates. The substrate holder, cleaned in an ultrasonic bath prior to mounting, is equipped with wiring for \textit{in-situ} 4-point measurements of the dc resistance of the growing film, using van der Pauw geometry. The electrical contacting to the sample is applied to the corners of the sample with silver paste. Via this arrangement we determine an average resistivity over the whole area of the film. After being mounted, the sapphire substrate was preheated at $80^\circ\mathrm{C}$ for one hour to reduce residual moisture. The substrate was then cooled with either liquid nitrogen or helium, depending on the targeted growth temperature, and the temperature was stabilized using a heater. A gradually increasing current, starting from 160~A was passed through the tungsten boat, to control the rate of evaporation of aluminum. Once the aluminum evaporation rate was stabilized at a desired value, the oxygen flow was adjusted, and the shutter over the substrate was opened to initiate the growth process. The film thickness was determined using a quartz crystal rate meter installed inside the evaporation chamber.

At substrate temperatures below 150~K, a significant challenge was ensuring proper film adhesion to the substrate after growth. This occurred because the substrate, being the coldest spot in the chamber during growth, attracted gases and particles that accumulated on its surface before the process started. To address this issue, we installed a cold finger that can be filled with liquid nitrogen prior to the growth process, making the cold finger the coldest spot in the chamber and thus redirecting the accumulation of condensing gases away from the substrate. This not only improved film adhesion to the substrate but also significantly reduced the chamber pressure from $1.2 \times 10^{-6} \, \text{mbar} \, \text{to} \, 2.5 \times 10^{-7} \, \text{mbar}$ at cryogenic temperatures. A copper heat shield was also installed to reduce heat input from the thermal radiation of the chamber walls. It was connected to the cryogenic tank with an adjustable clamp, allowing it to reach a lower temperature than the evaporation chamber walls. This setup shielded the sample holder, maintaining a lower temperature. Additionally, the opening at the bottom plate of the shield allowed only a small amount of evaporated material to directly reach the substrate holder, minimizing the temperature rise during growth \cite{DArneseThesis}.  

Two main parameters during the growth process discussed in this paper that affect the resistivity of the samples are oxygen flow in the chamber and rate of evaporation of aluminum. For room-temperature grown films, the oxygen flow was varied between 0.5 and 0.9~SCCM (standard cubic centimeters per minute). For films cooled with liquid nitrogen, we investigated samples grown at 150~K and 100~K, where the oxygen flow was varied between 0.1 and 1.0~SCCM. For the helium-cooled substrates, the temperature was set to 25~K, the lowest temperature we could stabilize during the growth process. For these films, the oxygen flow was varied between 0.2 and 0.6~SCCM, while higher oxygen flow at such low temperatures resulted in insulating films. The aluminum evaporation rate was adjusted according to different substrate temperatures. For room-temperature grown films, the rate was varied between 0.7 and 1.6~\AA/s. As the temperature was reduced, a higher rate was required to obtain a film of similar resistivity. For films cooled with liquid nitrogen, the aluminum evaporation rate was varied from 1.9 to 2.7~\AA/s at 150~K and from 1.8 to 3.1~\AA/s at 100~K. This trend continued for the helium-cooled substrates at 25~K, where the rate of evaporation was varied from 2.8 to 5~\AA/s. The oxygen flow rate was adjusted for each deposition and then held constant during the growth process. The evaporation rate was set to a desired value before the deposition and maintained relatively constant with small fluctuations. This careful adjustment of oxygen flow and rate of evaporation of aluminum allowed for the consistent deposition of films with comparable resistivities across the different growth temperatures, thereby facilitating a controlled study of their superconducting properties and tuning the superconducting dome \cite{Deshpande2025}. 

\begin{figure}[tb]
\centering
\includegraphics[width=12.5cm]{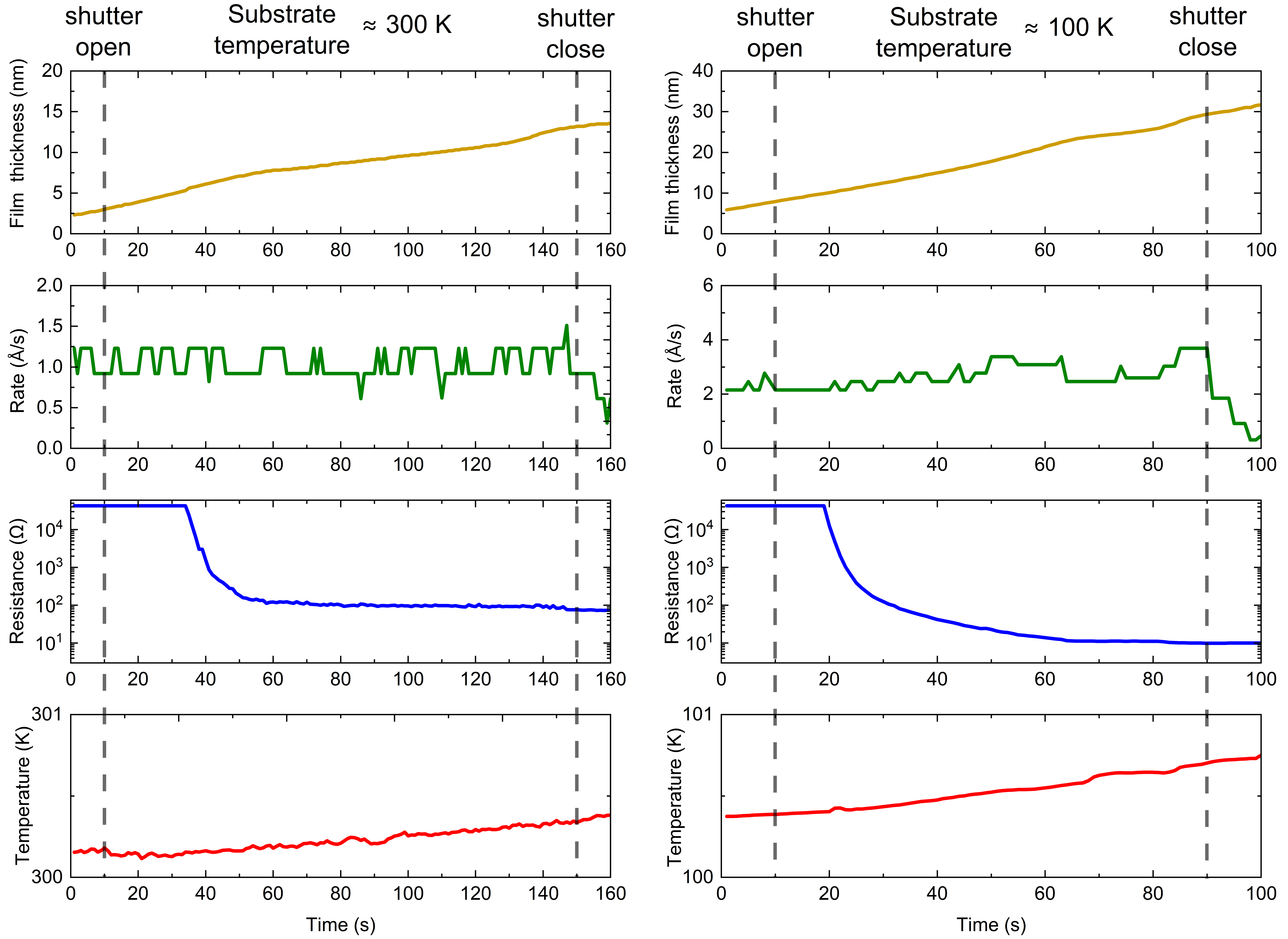}
\caption{Time evolution of deposition parameters during the thin film deposition at substrate temperatures of 300~K (left panel) and 100~K (right panel). Each panel (from top) shows film thickness, rate of evaporation, \textit{in-situ} resistance and substrate temperature over time. The dashed vertical lines indicate the opening and closing of the shutter, marking the start and end of the deposition process.}
\label{Fig:parameters stack}
\end{figure}

Figure \ref{Fig:parameters stack} shows the typical growth process and the time evolution of various parameters during growth at two different substrate temperatures, 300~K and 100~K, with film thicknesses of 10~nm and 20~nm, respectively. Once the rate is stabilized the shutter is opened, and closed again when the desired thickness is achieved. As can be seen from the upper panels in Fig. \ref{Fig:parameters stack}, the film thickness increases almost linearly over time, influenced by the stability of the evaporation rate during the growth process. The deposition rate shows typical fluctuations, which are expected in manual processes where current adjustments are needed. At lower substrate temperatures, the need for a higher evaporation rate makes stabilization more challenging, leading to increased rate fluctuations. Initially, the measured dc resistance is  very high and limited by our instrumentation, but as the film reaches a critical thickness of 3–5~nm, a conducting path is established between all four probes of the dc measurement, causing the resistance to drop sharply as the film grows. As seen in the lower panels of Fig. \ref{Fig:parameters stack}, the temperature changes very slightly throughout the process. 

After the growth process, the samples were carefully removed from the evaporator to the ambient lab atmosphere and then contacted again for temperature-dependent resistance measurements, performed in a separate $^4$He-cooled bath cryostat \cite{Rausch2018}. For all samples, the temperature-dependent dc resistivity was obtained from the 4-point dc resistance characterization in the cryostat and the thickness as determined during growth with the quartz rate meter. We focus on two quantities that characterize the electronic properties of each sample: the normal-state resistivity $\rho_{dc}$ at 5~K and $T_{c}$, with the latter defined as the temperature in the superconducting transition where the measured resistivity amounts to half of this normal-state value (50\% criterion)  \cite{Deshpande2025}. The samples were initially grown with varying thicknesses at different substrate temperatures under the assumption that thickness would not strongly affect the superconducting properties of granular aluminum. However, it was later realized that the thickness of the films does, in fact, influence the $T_{c}$ \cite{Deshpande2025}.

\section{Results and Discussion}\label{sec3}

\begin{figure}[tb]
\centering
\includegraphics[width=12.5cm]{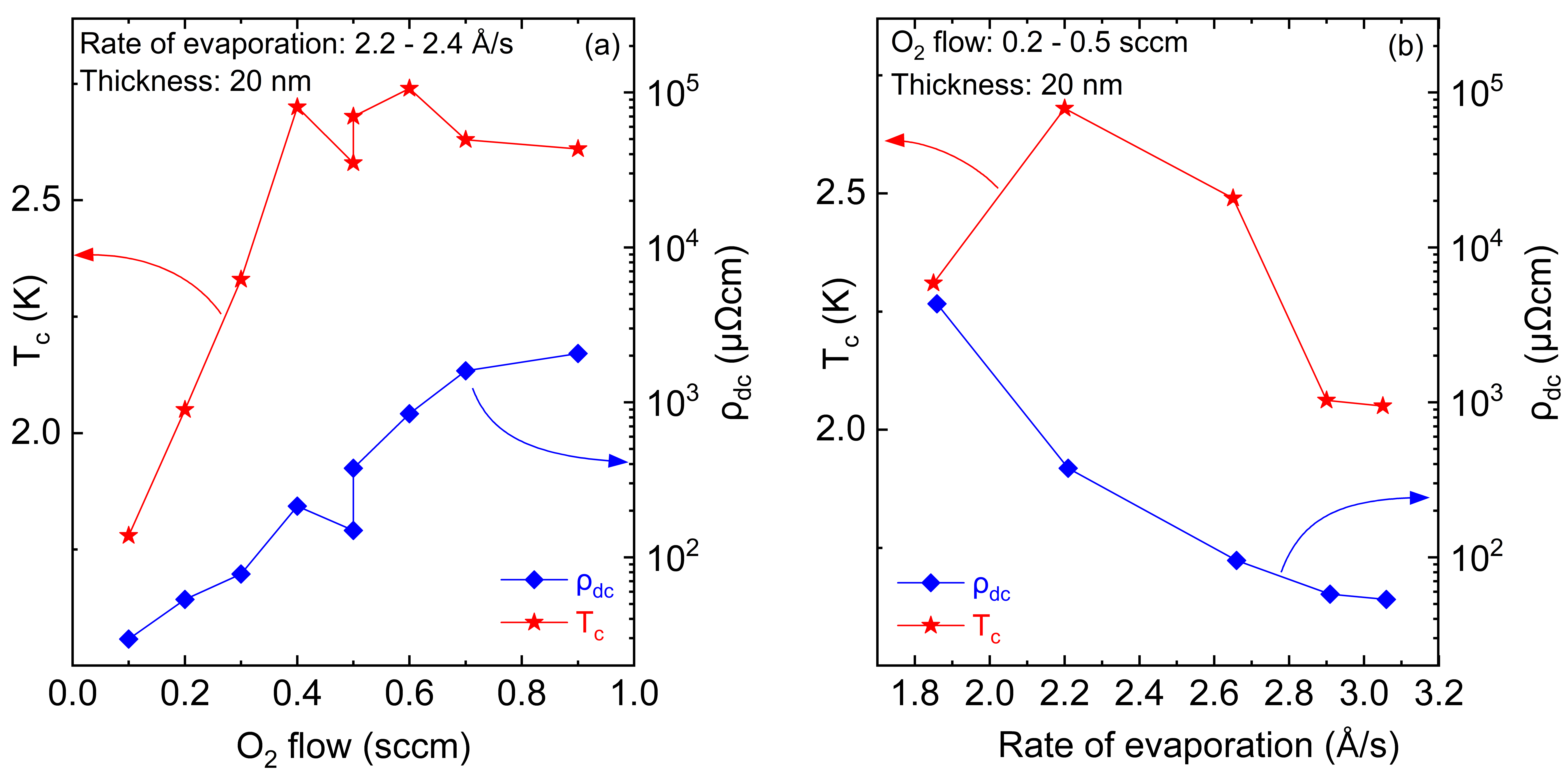}
\caption{Dependence of superconducting critical temperature ($T_{c}$) and normal-state resistivity ($\rho_{dc}$) on (a) oxygen flow rate and (b) rate of evaporation, at a substrate temperature of 100~K.}
\label{Fig:100K}
\end{figure}

The effect of the evaporation rate and oxygen flow in the chamber on the $\rho_{dc}$ and $T_{c}$ is shown in Fig. \ref{Fig:100K}, where 20~nm thick films were grown at a substrate temperature of 100~K. For the samples presented in Fig. \ref{Fig:100K}(a), the oxygen flow was then gradually increased from 0.1 to 0.9 SCCM for different films, while maintaining the evaporation rate within the range of 2.2-2.4 Å/s. We observe that the $\rho_{dc}$ increases with oxygen flow, as expected. 

\begin{figure}[tb]
\centering
\includegraphics[width=12.5cm]{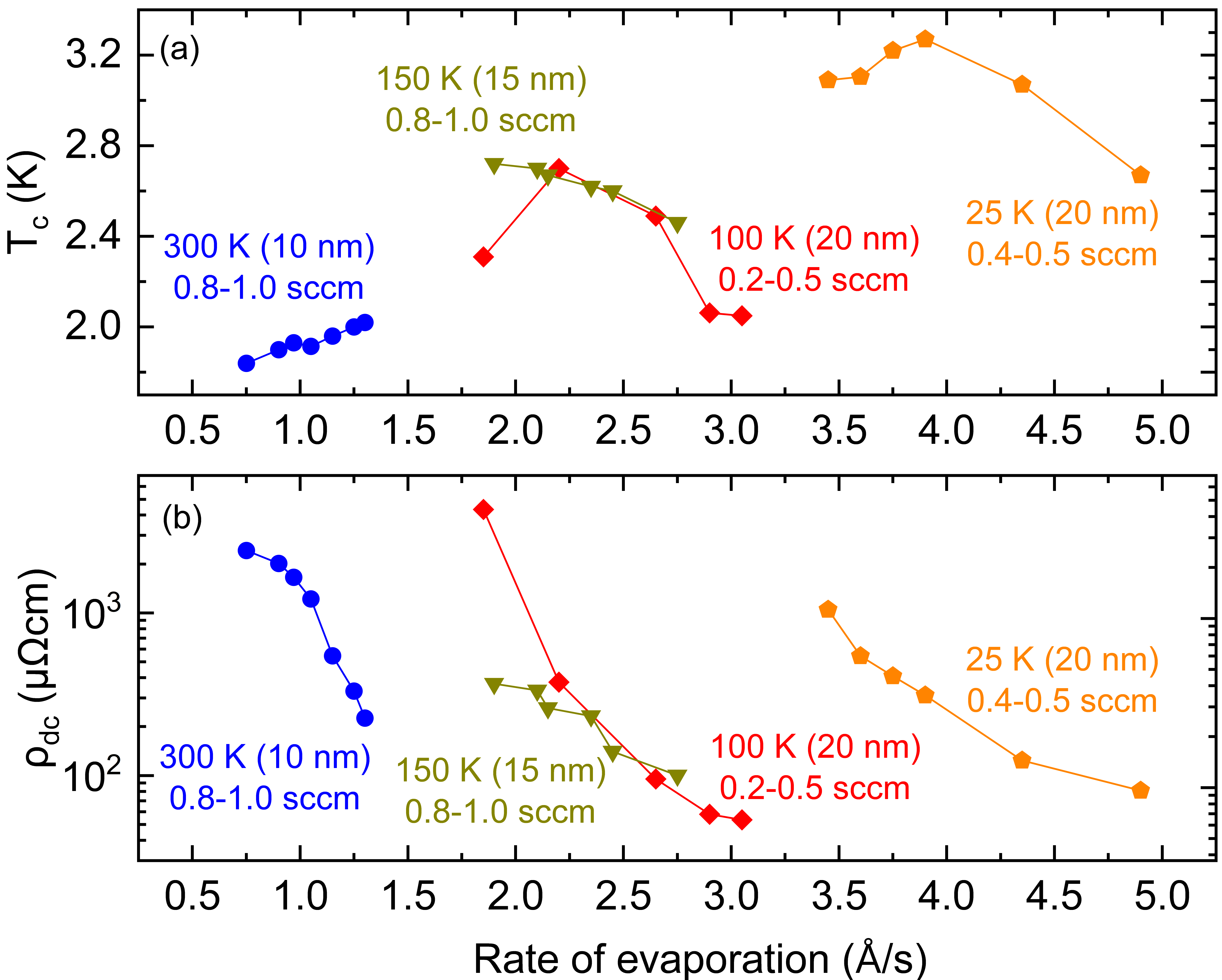}
\caption{Influence of the rate of evaporation on the (a) superconducting critical temperature ($T_{c}$) and  (b) normal-state resistivity ($\rho_{dc}$) for films deposited at different substrate temperatures. Each curve is labeled with the respective substrate temperature, film thickness, and oxygen flow rates.}
\label{Fig:rate_stack}
\end{figure}

For the data in Fig. \ref{Fig:100K}(b), a low oxygen flow range was set between 0.2 and 0.5~SCCM, and the rate of evaporation was varied from 1.8 to 3.1~\AA/s. The evaporation rate here is averaged over the entire growth process. We observe that as the evaporation rate increases, more aluminum is deposited relative to the oxygen incorporated in the films, leading to a decrease in resistivity. With resistivity spanning a broad range from 50 to 4000~$\mu\Omega$cm, the $T_{c}$ dependence on evaporation rate forms a characteristic superconducting dome shape. In Fig. \ref{Fig:100K}(a), two films were grown at 0.5~SCCM, with one having a slightly higher evaporation rate within the given range. This higher rate resulted in a dip in resistivity and a corresponding decrease in $T_{c}$.

Along with the deposition parameters such as aluminum evaporation rate and oxygen flow, the substrate temperature during growth plays a huge role in the growth process and the superconducting properties of the granular aluminum films \cite{Deshpande2025, Deutscher1973_JVST}. In Fig. \ref{Fig:rate_stack}(b), we plot $\rho_{dc}$ versus the rate of evaporation at different substrate temperatures. The gradual decrease in $\rho_{dc}$ with increasing evaporation rate, as shown in Fig. \ref{Fig:100K}(b), is expected. However, it is important to note that at lower substrate temperatures, achieving a film with similar resistivity requires a higher evaporation rate and lower oxygen flow. As we cool the substrate down, the surface mobility of aluminum atoms onto the sapphire substrate is significantly reduced, affecting the film’s growth dynamics and thus aluminum grain and oxide formation. To ensure proper nucleation and grain growth, and to achieve films with similar resistivity, a higher evaporation rate is required. Thus, in Fig. \ref{Fig:rate_stack}(b), at lower substrate temperatures like 100~K and 25~K, a reduced oxygen flow of less than 0.5~SCCM was used, whereas at higher temperatures, an oxygen flow of 0.8–1.0~SCCM was required. Fig. \ref{Fig:rate_stack}(a) shows how the superconducting transition temperature $T_c$ varies with the evaporation rate, at different substrate temperatures. Similar to Fig. \ref{Fig:100K}(b), the $T_c$ follows a dome-shaped dependence, where an optimal evaporation rate yields the highest $T_c$, while deviations from this rate, at both lower and higher values, result in a decrease in $T_c$. 
The balance of oxygen flow and evaporation rate at low temperatures ensures sufficient grain connectivity, which is critical for the particular macroscopic superconducting properties of granular aluminum that are widely tunable by proper growth conditions.

\section{Conclusions}\label{sec4}

We have examined the influence of oxygen flow, aluminum evaporation rate, and substrate temperature on the superconducting properties of granular aluminum films. Our findings show that the superconducting transition temperature and resistivity are significantly impacted by these growth parameters. This suggests that fine-tuning the deposition conditions is crucial for optimizing $T_{c}$, especially as deviations from the optimal rates and flows lead to significant variations in the superconducting properties.

Granular aluminum has often been viewed as a material with relatively  robust superconducting properties, despite variations in fabrication conditions. However, our study highlights a more intricate scenario. For example, the need for higher evaporation rates at lower temperatures, along with reduced oxygen flow to maintain proper grain connectivity, clearly indicates that detailed tuning of the growth parameters is needed to obtain films with the desired superconducting properties.

These results indicate that granular aluminum’s superconducting behavior is more complex than previously thought, especially when considering applications that rely on its high kinetic inductance. Our experience shows that it is more challenging to fully reproduce proper settings for the growth of samples with high dc resistivity and thus high kinetic inductance compared to samples with lower dc resistivity. The interplay between oxygen flow, evaporation rate, and the previously rarely explored parameter substrate temperature opens up avenues for material optimization, with the potential for further improvements in device performance in the fields of superconducting circuits and quantum technologies.

\section*{Acknowledgments}

We thank Gabriele Untereiner for support with sample preparation, Alessandro D’Arnese for initial work on the evaporator, and Nimrod Bachar, Ameya Nambisan, and Ioan Pop for helpful discussions.
This work was partially supported by the Deutsche Forschungsgemeinschaft DFG [grant SCHE 1580/6-1 / DR 228/57-1] and by the Baden-W\"urttemberg Stiftung [grant QT-10 (QEDHiNet)].

\bibliographystyle{elsarticle-num} 
\bibliography{refs_GrowthGranularAl_2025-01-07.bib}

\end{document}